\documentclass[sts]{imsart}

\RequirePackage[OT1]{fontenc}
\usepackage{amsthm,amsmath}

\usepackage{apacite,natbib}
\usepackage{colortbl}
\usepackage{float}
\usepackage[caption=false]{subfig}
\usepackage{multirow}
\RequirePackage[colorlinks,citecolor=blue,urlcolor=blue]{hyperref}
\usepackage{graphicx}
\usepackage{mathptmx}
\usepackage{epstopdf}

\usepackage{multirow}
\graphicspath{ {images/} }

\usepackage{lipsum}
\usepackage{tree-dvips}
\usepackage{xcolor}

\usepackage{float}
\usepackage{cases}
\newcommand\norm[1]{\left\lVert#1\right\rVert}

\def\tr{\mbox{tr}}

\startlocaldefs
\numberwithin{equation}{section}
\theoremstyle{plain}

\endlocaldefs

\begin{document}
\sloppy

\begin{frontmatter}
\title{High-dimensional covariance matrix regularization using informative targets}
\runtitle{Covariance matrix regularization using informative targets}

\begin{aug}
\author{\fnms{Atiq} \snm{Ur Rehman}\thanksref{t1}\ead[label=e1]{atiq.msst138@iiu.edu.pk}} \and
\author{\fnms{Muhammad} \snm{Farooq}\thanksref{t2}\ead[label=e2]{inshahullah03@yahoo.com}}


\thankstext{t1}{To whom correspondence should be addressed.}
\thankstext{t1}{Department of Mathematics and Statistics, Faculty of Basic and Applied Sciences, International Islamic University, Islamabad, Pakistan. Email: atiq.msst138@iiu.edu.pk}
\thankstext{t2}{Department of Statistics, University of Peshawar, Peshawar, Khyber Pakhtunkhwa,  Pakistan. Email: m.farooq@uop.edu.pk}
\runauthor{A.U. Rehman, and  M. Farooq}

\address{}
\end{aug}

\begin{abstract}
The sample covariance matrix becomes non-invertible in high-dimensional settings, making classical multivariate statistical methods inapplicable. Various regularization techniques address this issue by imposing a structured target matrix to improve stability and invertibility. While diagonal matrices are commonly used as targets due to their simplicity, more informative target matrices can enhance performance. This paper explores the use of such targets and estimates the underlying correlation parameter using maximum likelihood. The proposed method is analytically straightforward, computationally efficient, and more accurate than recent regularization techniques when targets are correctly specified. Its effectiveness is demonstrated through extensive simulations and a real-world application.
 


\end{abstract}


\end{frontmatter}

\section{Introduction}
\label{section1}

High-dimensional datasets, where the number of variables, $p$, is greater than the sample size, $n$, are becoming increasingly common in many fields, including genetics \citep[for example, see][]{eisen1998cluster,tamayo1999interpreting,beerenwinkel2007genetic}, brain imaging, and climate data. Publicly available gene expression data often contain 5,000--10,000 genes and fewer than 100 samples, and both numbers are expected to grow over time \citep{dudoit2002comparison}. In such high-dimensional settings, the sample covariance matrix---whether the maximum likelihood estimator or its unbiased version---performs poorly and does not approximate the true covariance matrix well, even when $p$ is comparable to $n$. This is because the sample covariance matrix contains estimation errors, leading to overestimated large eigenvalues and underestimated small eigenvalues. Moreover, the inverse covariance matrix is fundamental to classical multivariate methods such as regression, linear and quadratic discriminant analysis, Gaussian graphical models, and Mahalanobis distance. However, when $p > n$, the sample covariance matrix loses full rank and becomes non-invertible, precluding the use of these methods.

Several approaches have been proposed to regularize the covariance matrix estimation and approximate its inverse in high-dimensional settings. One of the earliest attempts is the Moore-Penrose generalized inverse \citep{penrose1955generalized}, which relies on singular value decomposition (SVD) to approximate the inverse covariance matrix using only nonzero singular values and their corresponding singular vectors. Shrinkage estimation \citep{ledoit2003improved,ledoit2004honey} combines the sample covariance matrix with a positive definite target matrix, where the combination proportion (shrinkage intensity) is estimated by minimizing the Frobenius norm. \cite{schafer2005shrinkage} extended this approach by introducing different target matrices, including the identity matrix. Notably, ridge regression \citep{hoerl1970ridge} is a special case of shrinkage estimation. Another method, proposed by \cite{warton2008penalized}, applies a penalized likelihood function with a penalty parameter chosen through cross-validation. Additional regularization techniques, such as LASSO \citep{friedman2008sparse}, induce sparsity in the inverse covariance matrix by penalizing the Gaussian likelihood function. \cite{pourahmadi2013high} provides a comprehensive review of well-known covariance matrix regularization methods.

The choice of the target matrix significantly influences the accuracy of a shrinkage estimator \citep[see][]{schafer2005shrinkage}. When correctly specified, the target matrix can greatly improve estimation accuracy. \cite{ledoit2004honey} emphasized two key properties of a good target: (1) it should have a small number of parameters to estimate, and (2) it should capture important structural characteristics of the true covariance matrix. The identity matrix, for instance, is a parameter-free target that satisfies the first condition but may fail to capture the true covariance structure. In practice, selecting an appropriate target often requires domain knowledge or exploratory data analysis.

In high-dimensional data analysis, the primary objective is often to make computations feasible rather than to improve estimation accuracy. The identity matrix is widely used as a target due to its simplicity and the common assumption of sparsity, which is reasonable in many high-dimensional scenarios. For example, the R package *corpcor* \citep{schaefer2013corpcor} defaults to using the identity matrix as a target. However, in certain cases, using the identity matrix may be inappropriate, and more informative target matrices can improve performance.

This paper aims to demonstrate situations where using the identity matrix as a target is suboptimal and to explore the benefits of more informative targets. Additionally, we propose a novel method for estimating the penalty parameter based on the Gaussian likelihood function. The same likelihood function is used to estimate the correlation parameter, which plays a crucial role in both autoregressive and exchangeable covariance structures.

The paper is organized as follows: Our method is described in Sections \ref{section2} and \ref{section3}, followed by applications to simulated and real-world data in Sections \ref{section4} and \ref{section5}, respectively. Conclusions are presented in Section \ref{section6}.

\section{Estimation of regularization parameter using normal likelihood} 
\label{section2}

Let the real-valued observations are contained in $X=(x_1,\hdots,x_p)$, 
where $x_i$ is a $p$-dimensional sample realization made independently over $n$ objects.
Let $\boldsymbol{\mu}$ denote the mean vector and $\boldsymbol{\Sigma}$ the covariance matrix. Assume also that the data in $X$ is centred so that $\mu=0$, and thus the Gaussian log-likelihood function is 
\begin{equation}
\log L(\boldsymbol{X};\boldsymbol{\Sigma})= constant - \frac{n}{2} \log\left|\boldsymbol{\Sigma}\right|-\frac{1}{2} tr(\boldsymbol{X^t \Sigma^{-1} X}), 
\label{equ17} 
\end{equation}
where the first term on the right hand side does not depend on $\boldsymbol{\Sigma}$. The maximum likelihood estimate of $\boldsymbol{\Sigma}$ is 
$$
\boldsymbol{S}=\frac{1}{n} X^tX
$$ 
and its unbiased version is given by
$$
\widehat{\boldsymbol{\Sigma}}_u=\frac{1}{n-1} X^tX.
$$
In high-dimensional applications the sample estimator (maximum likelihood estimate or its unbiased version) of the covariance matrix is not invertible. Regularization procedures make the estimated covariance invertible and more stable through incorporating additional information about the structure of the covariance matrix in the from of a target. For example, it may be reasonable to shrink the elements of a covariance matrix towards a diagonal matrix assuming sparsity, which is a sensible assumption in high-dimensional setting. The regularized estimate (a.k.a shrinkage estimate) takes the following form:
\begin{equation}
\boldsymbol{\hat{\Sigma}}_{\kappa} = \boldsymbol{S} + \kappa \boldsymbol{T},
\label{equshr}
\end{equation} 
where $\boldsymbol{T}$ is a positive definite (possibly informative) target matrix and $\kappa > 0$ is the regularization parameter. The expression in \ref{equshr} incorporates additional information we might have about the structure of the covariance matrix through the target $\boldsymbol{T}$. It is important for the parameter $\kappa$ to have the following properties (which we have explained in this section later):
\begin{itemize}
\item For fixed $p$ as we increase $n$, $\kappa\to 0$. Consequently, $\boldsymbol{\hat{\Sigma}}_{\kappa}\to\boldsymbol{S}$.
\item For fixed $n$ as we increase $p$, $\kappa\to\infty$. Consequently, $\boldsymbol{\hat{\Sigma}}_{\kappa}\to\boldsymbol{T}$.
\end{itemize} 

Here we make use of the second property and assume that $n$ is very small compared to $p$. We also assume that $\boldsymbol{T}$ is informative (structurally correctly specified).
In this situation $\boldsymbol{\hat{\Sigma}}_{\kappa}$ becomes similar (if not equal) to $\boldsymbol{T}$, that is  
\begin{equation*}
\boldsymbol{T} \approx \boldsymbol{S} + \kappa \boldsymbol{T}
\end{equation*}
or 
\begin{equation}
\kappa \boldsymbol{T} \approx \boldsymbol{S}-\boldsymbol{T},
\label{equnshr}
\end{equation}
which can be exploited to find the value of $\kappa$ as we do later in this section.

The expression in \ref{equnshr} can also be obtained by assuming that some scaled version of $\boldsymbol{T}$ is the true covariance and replacing $\boldsymbol{\Sigma}$ by $(1+\kappa)\boldsymbol{T}$ in equation \ref{equ17}, which then becomes
\begin{equation}
 \log L(\boldsymbol{X};\boldsymbol{\Sigma}_{\kappa}) = Const - \frac{n}{2} \log\left|\boldsymbol{T} + \kappa \boldsymbol{T} \right|-\frac{1}{2}\boldsymbol{X^t \frac{1}{1+\kappa}\boldsymbol{T}^{-1} X}.
 \label{equlik}
\end{equation} 
Differentiating \ref{equlik} with respect to $\kappa$ we get
\begin{equation}
\frac{\partial}{\partial \kappa} \log L(\boldsymbol{X};\boldsymbol{\Sigma}_{\kappa}) = - \frac{n}{2} (\boldsymbol{T} + \kappa \boldsymbol{T})^{-1} \boldsymbol{T} + \frac{1}{2} (\boldsymbol{T} + \kappa \boldsymbol{T})^{-1} T (X^t  X) (\boldsymbol{T} + \kappa \boldsymbol{T})^{-1},
\label{equ18}
\end{equation}
where $\frac{\partial}{\partial \boldsymbol{\kappa}} \log |\boldsymbol{A}| = \frac{1}{|\boldsymbol{A}|} |\boldsymbol{A}| \boldsymbol{A}^{-t}$ and $\frac{\partial}{\partial \boldsymbol{\kappa}} \boldsymbol{A}^{-1} = -\boldsymbol{A}^{-1} \frac{\partial A}{\partial \boldsymbol{\kappa}} \boldsymbol{A}^{-1}$. Equating equation \ref{equ18} to zero leads to
\begin{equation}
 \kappa \boldsymbol{T} = \boldsymbol{S} - \boldsymbol{T},
 \label{flik}
\end{equation}
which is similar to the expression in \ref{equnshr}. This leads to
\begin{equation}
 \norm{\kappa \boldsymbol{T}}_{1} = \norm{\boldsymbol{S} - \boldsymbol{T}}_{1},
\label{equ19}
\end{equation}
where $\norm{.}_{1}$ denotes the sum of the absolute values of all entries in the matrix.
If target $\boldsymbol{T}$ is the identity matrix then the value of $\kappa$ becomes
\begin{equation}
\kappa = \frac{\norm{\boldsymbol{S} - \boldsymbol{I}}_{1}}{p}.
\label{ident-kappa}
\end{equation}
Using AR(1) covariance structure as a target the expression for estimating $\kappa$ becomes
\begin{equation}
\kappa = \frac{\norm{\boldsymbol{S} - \boldsymbol{T}}_{1}}{p + 2\sum_{k=1}^{p-1} kt^{p-k}},
\label{ar1-kappa}
\end{equation}
and if $\boldsymbol{T}$ is of exchangeable structure then
\begin{equation}
\kappa = \frac{\norm{\boldsymbol{S} - \boldsymbol{T}}_{1}}{p + p(p-1)t}.
\label{exch-kappa}
\end{equation}
It is important to highlight that for $t=0$, \ref{ident-kappa} is a special case of \ref{ar1-kappa} and \ref{exch-kappa}. In all three cases the value of $\hat{\kappa}$ is the weighted average of the element-wise absolute error. Under a correctly specified target, the weighted average of element-wise absolute error tends to decrease as we increase $n$ for any fixed value of $p$, since $\boldsymbol{S}$ is a consistent estimator of $\boldsymbol{\Sigma}$. Therefore, the larger error leads to larger values of $\hat{\kappa}$ (hence more reliance on $\boldsymbol{T}$), and vice versa if the error is small. 

In order to restrict the range of the regularization parameter between zero and one, it is more appropriate to use the correlation scale rather than the covariance scale in equation \ref{equ19}. The correlation matrix can be obtained as 
\begin{equation*}
\boldsymbol{\hat{R}} = \boldsymbol{S}_{d}^{-\frac{1}{2}} \boldsymbol{S} \boldsymbol{S}_{d}^{-\frac{1}{2}},
\end{equation*}
where $\boldsymbol{S_{d}}$ is the diagonal matrix with corresponding diagonal elements of $\boldsymbol{S}$ on the main diagonal. To get the regularization parameter at correlation scale we follow \cite{warton2008penalized} who regularized the correlation matrix (not the covariance matrix) and rescaled $\kappa$ as
 \begin{equation}
 \gamma = \frac{1}{1+\kappa},
 \end{equation}
where $\gamma \in [0,1]$. The corresponding regularized estimator of the correlation matrix can be obtained as
\begin{equation}
\hat{\boldsymbol{R}}_{\gamma} = \gamma \boldsymbol{\hat{R}}  + (1-\gamma) \boldsymbol{T}.
\end{equation}

\section{Estimation of Correlation parameter} 
\label{section3}

Unlike the identity matrix, the AR(1) and exchangeable targets rely on one additional parameter, which needs to be estimated. The AR(1) covariance is given by
\begin{equation}
     \sigma_{ij} = t^{|i-j|} \quad     \text{for} \quad  1 \le i,j \le p,
     \label{equ2}
\end{equation}
and the exchangeable covariance structure is given by
\begin{equation}
\sigma_{ij} = 
\begin{cases}
                                   1 & \text{when $i=j$} \\
                                   t & \text{when $i \neq j$} 
  \end{cases} \quad     \text{for} \quad  1 \le i,j \le p,
  \label{equ3}
\end{equation}
where $t$ is an unknown parameter and can be estimated from the data \citep{wang2003working}. We denote the covariance structures by  $\boldsymbol{\Sigma_{t}}$ to make its reliance on $t$ explicit. The estimate of $t$ is obtained by simply maximizing the $\log$-likelihood function of the multivariate normal distribution, that is
 \begin{equation}
 \frac{\partial}{\partial t} \log L(\boldsymbol{X};\boldsymbol{\Sigma_{t}}) = - \frac{n}{2} \frac{\partial}{\partial t} \log\left|\boldsymbol{\Sigma_{t}}\right|-\frac{1}{2} \tr \big ( \boldsymbol{X^t \frac{\partial}{\partial t} \Sigma_{t}^{-1} X} \big ).
\label{equ5}
\end{equation}
Now if $\boldsymbol{\Sigma_{t}}$ has AR(1) covariance structure then the determinant of $\boldsymbol{\Sigma_{t}}$ is $\boldsymbol{|\Sigma_{t}|} = (1- t^2)^{p-1} $. Differentiating $\log\left|\boldsymbol{\Sigma_{t}}\right|$ with respect to $t$ we have $\frac{\partial}{\partial t} \log |\Sigma_{t}|= \frac{-2(p-1)t(1-t^2)^{p-2}}{(1-t^2)^{p-1}}$.
The inverse of $\boldsymbol{\Sigma_{t}}$ is given by
\begin{equation*}
\Sigma_{t}^{-1}= \frac{1}{(1-t^2)}
  \begin{pmatrix}
    1 & -t & 0 & ... & 0 & 0 \\
    -t & 1+t^2 & -t & ... & 0 & 0 \\
    0 & -t & 1+t^2 & ... & 0 & 0 \\
    \vdots   & \vdots & \vdots &  \vdots &   \vdots & \vdots\\
    0 & 0 & 0 & ... & 1+t^2 & -t  \\
    0 & 0 & 0 & ... & -t & 1 \\
  \end{pmatrix}.
\end{equation*}
Differentiating $ \boldsymbol{\Sigma_{t}^{-1}}$ with respect to $t$ is obtain
   \begin{equation*}
\frac{\partial}{\partial t} \Sigma_{t}^{-1}= \frac{1}{(1-t^2)^2}
  \begin{pmatrix}
    2t & -(1+t^2) & 0 & \hdots & 0 & 0 \\
    -(1+t^2) & 4t & -(1+t^2) & \hdots & 0 & 0 \\
    \vdots & \vdots & \vdots & \vdots & \vdots & \vdots \\
    0 & 0 & 0 & \hdots & 4t & -(1+t^2)  \\
    0 & 0 & 0 & \hdots & -(1+t^2) & 2t \\
  \end{pmatrix}.  
  \end{equation*}
To find $\tr \big( X^t \frac{\partial}{\partial t} \Sigma_{t}^{-1} X \big )$ in equation \ref{equ5}, we can write $\boldsymbol{X^t X = n \Sigma}$, where $\boldsymbol{\Sigma}$ is true covariance matrix with entries $\sigma_{ij}, 1 \le i,j \le p$. then,
\begin{equation}
\begin{split}
n \, \tr \big ( \frac{\partial}{\partial t} \Sigma_{t}^{-1} \Sigma \big )= &\frac{1}{(1-t^2)^2} [ 2t(\sigma_{11} + 2\sigma_{22} + 2\sigma_{33} + ... + 2\sigma_{(p-1)(p-1)} + \sigma_{pp})]\\
& - t^2(\sigma_{12} + \sigma_{21} + \sigma_{23} + ... + \sigma_{(p-1)(p)} + \sigma_{p(p-1)})\\
& - (\sigma_{12} + \sigma_{21} + \sigma_{23} + ... + \sigma_{(p-1)(p)} + \sigma_{p(p-1)}).
\end{split}
\label{equ8}
\end{equation}
Since $\boldsymbol{\Sigma}$ is a symmetric matrix (that is,  $\sigma_{ij}=\sigma_{ji}$) and its diagonal entries are all unity, we can write equation \ref{equ8} as

\begin{equation}
n \, \tr \big ( \frac{\partial}{\partial t} \Sigma_{t}^{-1} \Sigma \big ) =\frac{2}{(1+t^2)^2} [t(p-1) - (t^2 + 1) \sum_{i=2}^{p} \sigma_{(i-1)i}].
\label{equ9}
\end{equation}
Substituting equation $\frac{\partial}{\partial t} \log |\Sigma_{t}|$ and $n \, \tr \big ( \frac{\partial}{\partial t} \Sigma_{t}^{-1} \Sigma \big )$ in equation \ref{equ5} and equating it to zero leads to
\begin{equation}
\hat{t} = \frac{\sum_{i=2}^{p} \hat{\sigma}_{(i-1)i}}{p-1}.
\label{equ11}
\end{equation}
Similarly, if $\boldsymbol{\Sigma_{t}}$ has exchangeable structure then $|\Sigma_{t}| = (1-t)^{p-1} \lbrace 1+(p-1)t \rbrace$. Differentiating $\log |\Sigma_{t}|$ with respect to $t$ gives
 \begin{equation}
    \frac{\partial}{\partial t} \log|\Sigma_{t}| = \frac{(1-t)^{p-1} (p-1) + \lbrace 1+(p-1)t \rbrace (p-1)(1-t)^{p-2}}{(1-t)^{p-1} \lbrace 1+(p-1)t \rbrace}.
    \label{equ12}
   \end{equation}
In this case the inverse of $\Sigma_{t}$ is
 \begin{equation*}
 \begin{split}
 \Sigma_{t}^{-1}=& \frac{1}{(1-t) \lbrace 1+(p-1)t \rbrace}
  \begin{pmatrix}
    1+t & -t & \hdots & -t  \\
    -t & 1+t & \hdots & -t  \\
    \vdots & \vdots & \vdots & \vdots \\
    -t & -t  & \hdots & 1+t
  \end{pmatrix}\\
&=\frac{1}{(1-t)} \big [\textbf{I} - \frac{t}{ \lbrace 1+(p-1)t \rbrace} \textbf{J} \big ],
  \end{split}
 \end{equation*}
where $\textbf{I}$ is the identity matrix and $\textbf{J}$ is the unit matrix. Differentiating $\Sigma_{t}^{-1}$ with respect to $t$ we get 
\begin{equation}
\frac{\partial}{\partial t} \Sigma_{t}^{-1} = \frac{1}{(1-t)^2} \textbf{I} - \Big [ \frac{1}{ \lbrace 1+(p-1)t \rbrace^2 (1-t)} + \frac{t}{\lbrace 1+(p-1)t \rbrace(1-t)^2} \Big ] \textbf{J}.
\label{equ14}
\end{equation}
Then, 
\begin{equation}
\begin{split}
\tr(\frac{\partial}{\partial t} \Sigma_{t}^{-1} \Sigma) =& \frac{p}{(1-t)^2} - p \big [ \frac{1}{\lbrace1+(p-1)t\rbrace^2 (1-t)} + \frac{t}{\lbrace1+(p-1)t\rbrace(1-t)^2} \big ]\\
& - \big [ \frac{1}{\lbrace1+(p-1)t\rbrace^2 (1-t)} + \frac{t}{\lbrace1+(p-1)t\rbrace(1-t)^2} \big ] \sum_{i \neq j}^{p} \sigma_{ij}.
 \end{split}   
  \label{equ15}
\end{equation}
Substituting the values of $\frac{\partial}{\partial t} \log|\Sigma_{t}|$ and $n \, \tr(\frac{\partial}{\partial t} \Sigma_{t}^{-1} \Sigma)$, respectively, from \ref{equ12} and \ref{equ15} into equation \ref{equ5} and equating it to zero leads to
\begin{equation}
\hat{t} = \frac{\sum_{i \neq j}^{p} \hat{\sigma}_{ij}}{p(p-1)}.
\label{equ16}
\end{equation}

\section{Simulation study}
\label{section4}

In this section, we present resutls of an extensive simulation study conducted to numerically demonstrate the performance of the proposed method in comparison with the shrinkage method proposed by \cite{schafer2005shrinkage} and implemented in R package ``corpcor'' \citep{schaefer2013corpcor}. To examine the behaviour of the proposed method under different simulation settings, we generate various datasets taking into account a number of different factors. These factors include varying sample sizes, numbers of variables, and covariance structures.

We draw a sample of size $n$ from a $p$-variate normal distribution with mean vector, $\boldsymbol{\mu=0}$, and covariance matrix, $\boldsymbol{\Sigma}$. In order to evaluate the performance in a variety of situations, we consider four different covariance structures for $\boldsymbol{\Sigma}$: AR(1), exchangeable, random covariance structure  \citep[see the algorithm in][]{schafer2004empirical} and an identity matrix.

\subsection{Empirical properties of $\boldsymbol{\hat{\gamma}}$}

\begin{figure}[H]
\begin{center}$
\begin{array}{ll}
\includegraphics[scale=0.33]{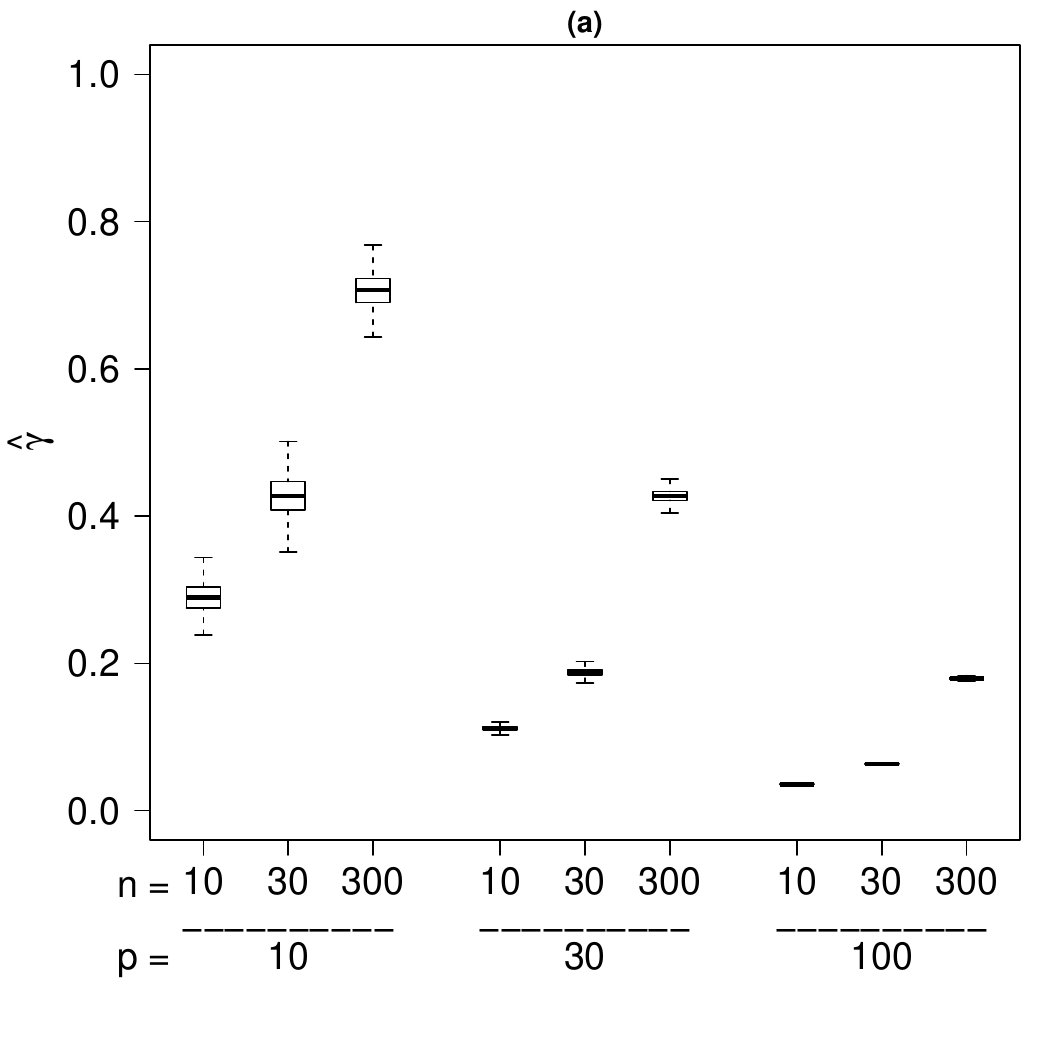}\\
\includegraphics[scale=0.33]{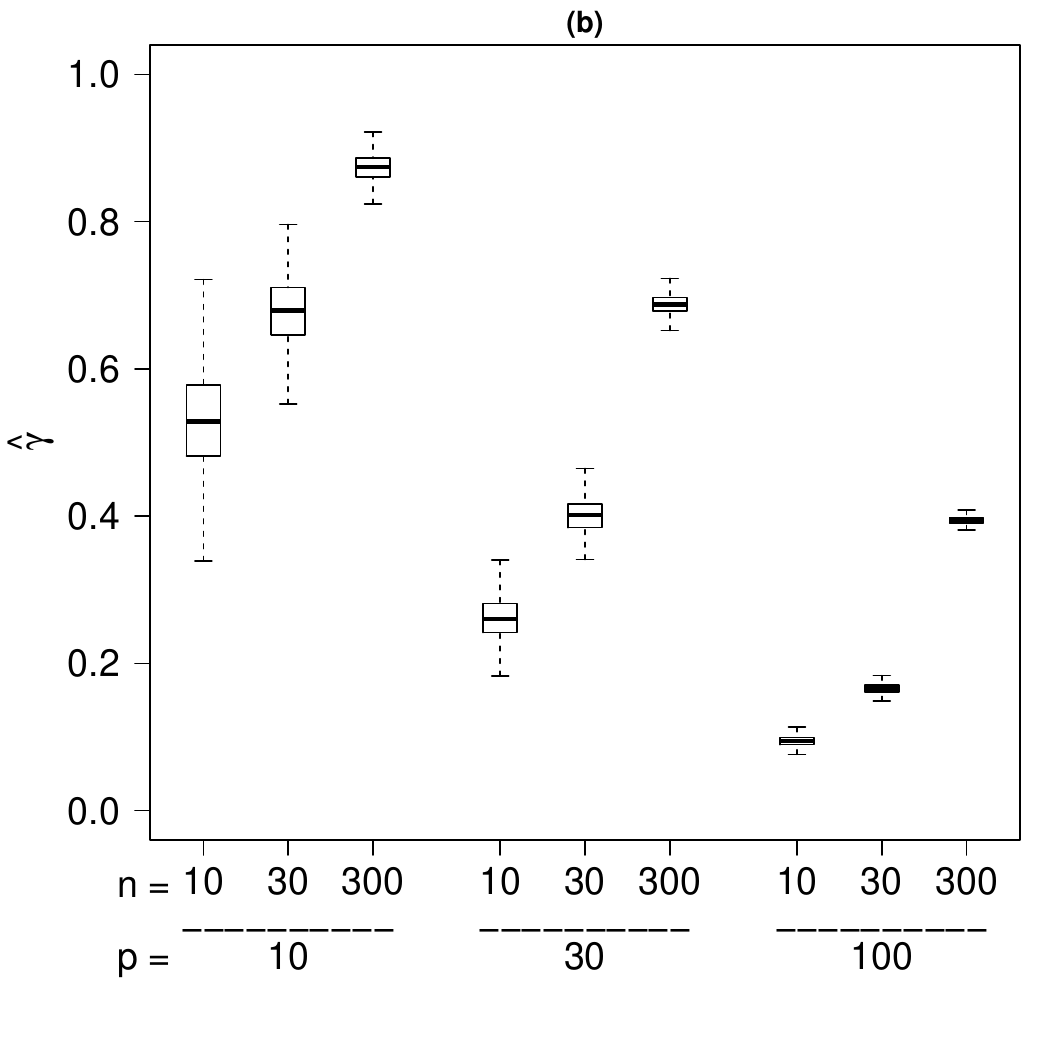}\\
\includegraphics[scale=0.33]{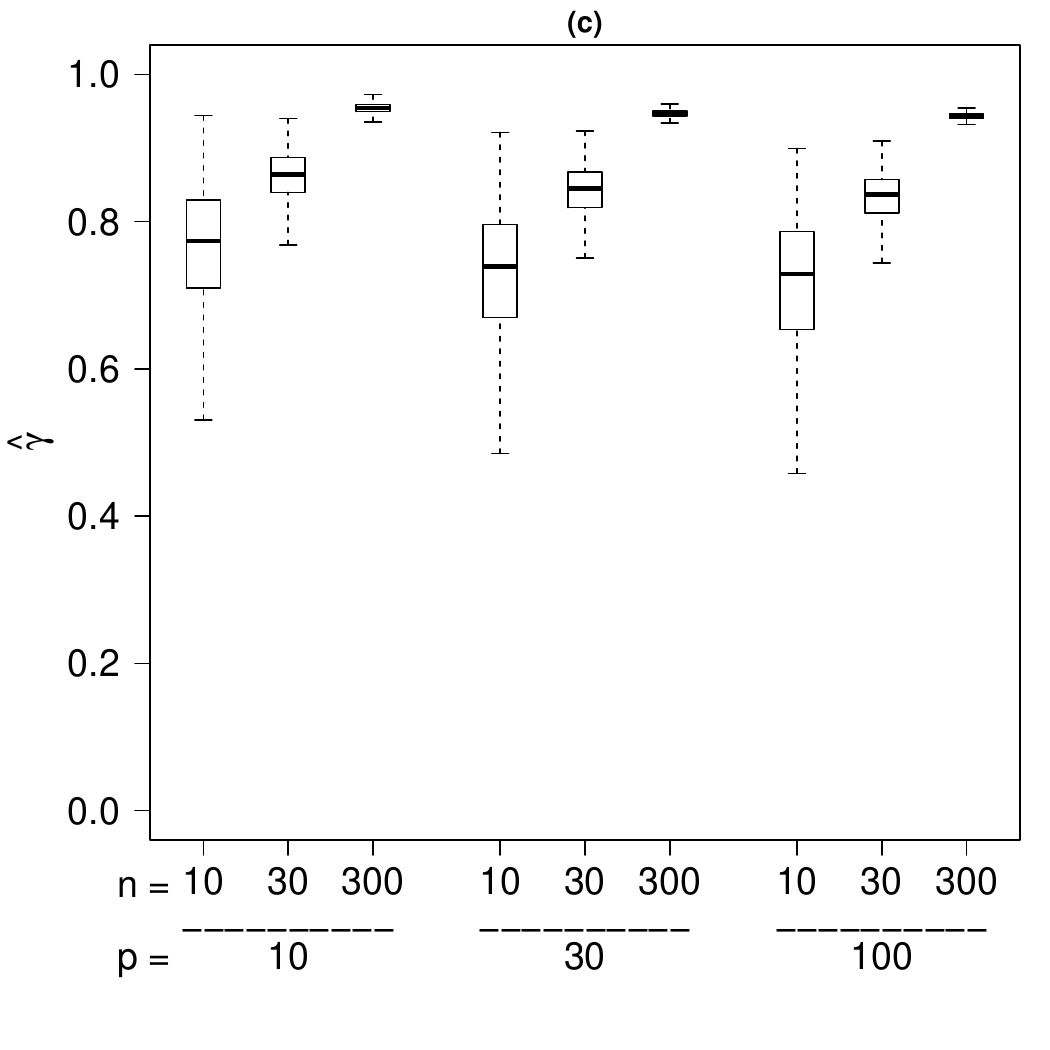}
\end{array}$
\end{center}
\caption{Distribution of $\hat{\gamma}$ values obtained in 1000 simulations from multivariate normal distribution with $\boldsymbol{\mu=0}$ and covariance matrix: (a) identity (b) AR(1) with $t=.5$ (c) exchangeable with $t=.5$. The target matrices are correctly specified in all three cases: (a) identity (b) AR(1) and (c) exchangeable. The results are shown for $n\in \lbrace10,30,300\rbrace$ and $p\in \lbrace10,30,100\rbrace$.}
\label{Fig3.1}
\end{figure}

Here we examine the properties of $\hat{\gamma}$ under different simulation settings. The values of $\hat{\gamma}$ exhibit some interesting properties in simulations as summarized in Figure \ref{Fig3.1}. First, for a fixed $p$, as $n$ increases, $\hat{\gamma}$ on the average increases, which indicates that as more data become available the sample covariance is less penalized. In other words the regularized estimator, $\boldsymbol{\Sigma}_{\gamma}$, is a consistent estimator of the true covariance matrix. Second, for a fixed $n$, as $p$ increases, $\hat{\gamma}$ on the average decreases. This translates to increased reliance of $\boldsymbol{\Sigma}_{\gamma}$ on the information provided (in the form of a target matrix) with increasing instability in $\boldsymbol{S}$. 

The $\hat{\gamma}$ values are not consistent across different covariance structures considered here. This is mostly because of the targets specified,which we specify correctly in all three cases; by ~~correctly", we mean when both true covariance matrix and target have same structure. In case of the identity matrix as a target no parameter is needed to be estimated from the data. Thus with increasing $n$, the sample covariance becomes more accurate, which results in decreased reliance on the target matrix. In case of AR(1) and exchangeable structures, the targets also depend on the data through the estimate of the parameter $t$. Estimating the parameter $t$ accurately translates to accurate estimation of the entire target matrix. Therefore, with increasing $n$ both the sample covariance and the targets become closer to the true covariance. This is more obvious in the case of exchangeable covariance.

Note also that the variability in $\hat{\gamma}$ tend to decrease as $n$ increases, for a fixed value of $p$.

\subsection{Comparative performance of $\widehat{\boldsymbol{\Sigma}}_{\gamma}$}

We used the sum of absolute errors in eigenvalues to compare the accuracy of the proposed method with the maximum likelihood estimate and the shrinkage method of \cite{schafer2005shrinkage}. We showed the results for $n=50$ and $p\in\lbrace 30, 50, 100\rbrace$ to demonstrate the effect of increasing number of variables for a fixed value of $n$. For AR(1) and exchangeable covariance structures, we shrunk the estimated covariance matrix towards the correct targets that are, respectively, AR(1) and exchangeable. We also examined the performance in the cases where target matrix is incorrectly specified as identity matrix while the true covariance matrix is AR(1) or exchangeable. However, for random covariance structure we used only identity matrix as a target, which although incorrect, has been used extensively as a shrinkage target to regularize the covariance matrix \citep[see, for example,][]{ledoit2003improved, schafer2005shrinkage}. Note that although we have conducted the experiments for a range of values of $t$, we show here the results only for $t=0.5$ for both AR(1) and exchangeable structures (due to consistency of the results for different values of $t$). Similarly, for random covariance structure we showed results for a covariance matrix whose inverse contains 30\% of the off-diagonal positions as being non-zero. 

\begin{figure}[H]
\begin{center}$
\begin{array}{ll}
\includegraphics[scale=0.33]{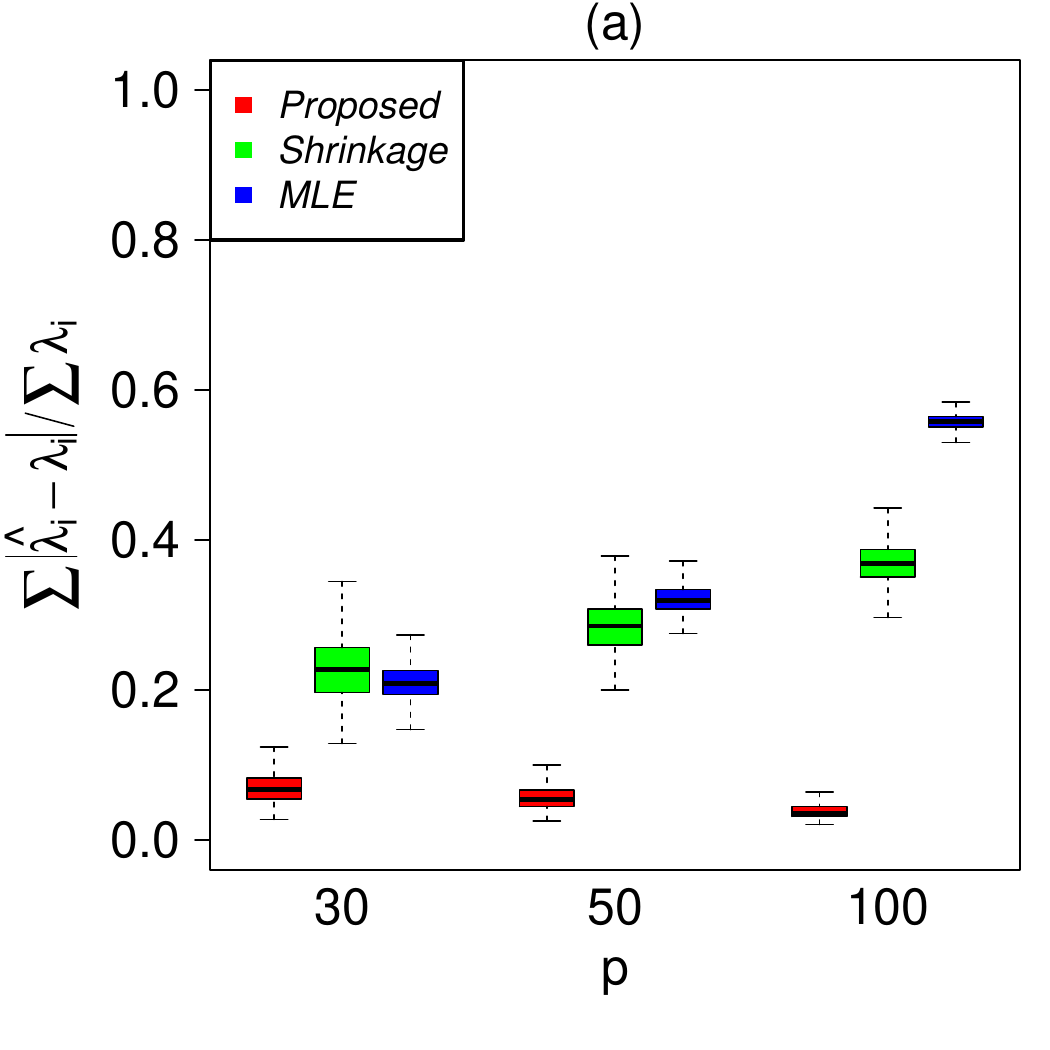}
\includegraphics[scale=0.33]{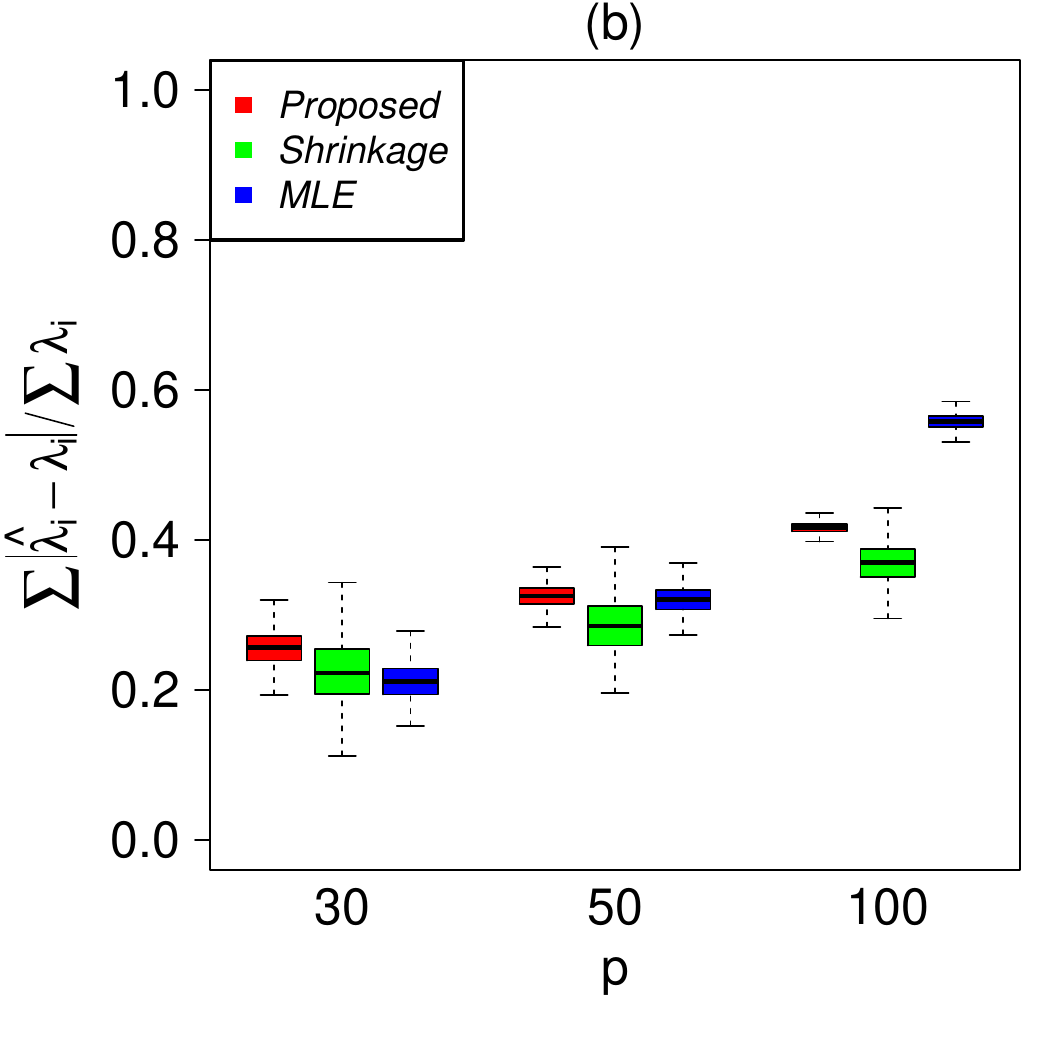}
\end{array}$
\end{center}
\begin{center}$
\begin{array}{ll}
\includegraphics[scale=0.33]{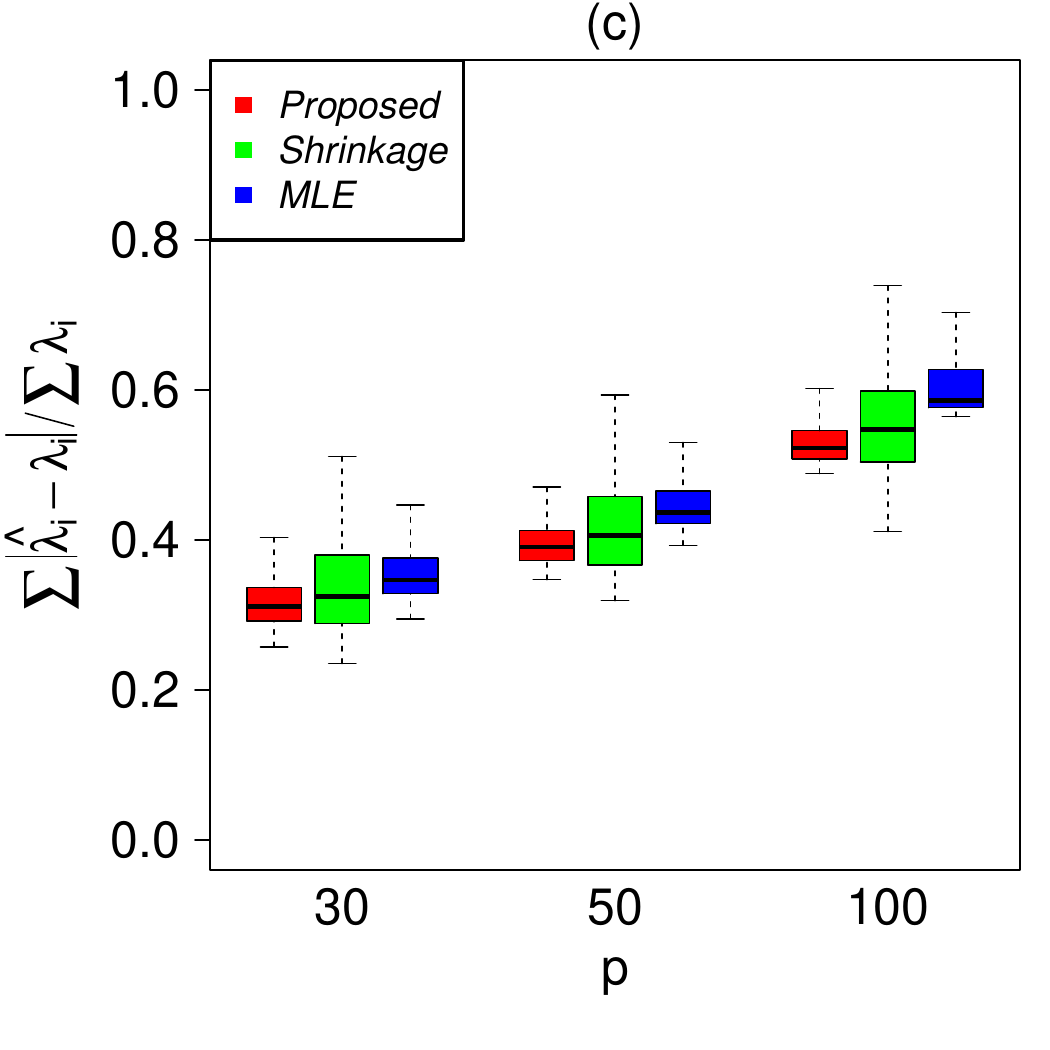}
\includegraphics[scale=0.33]{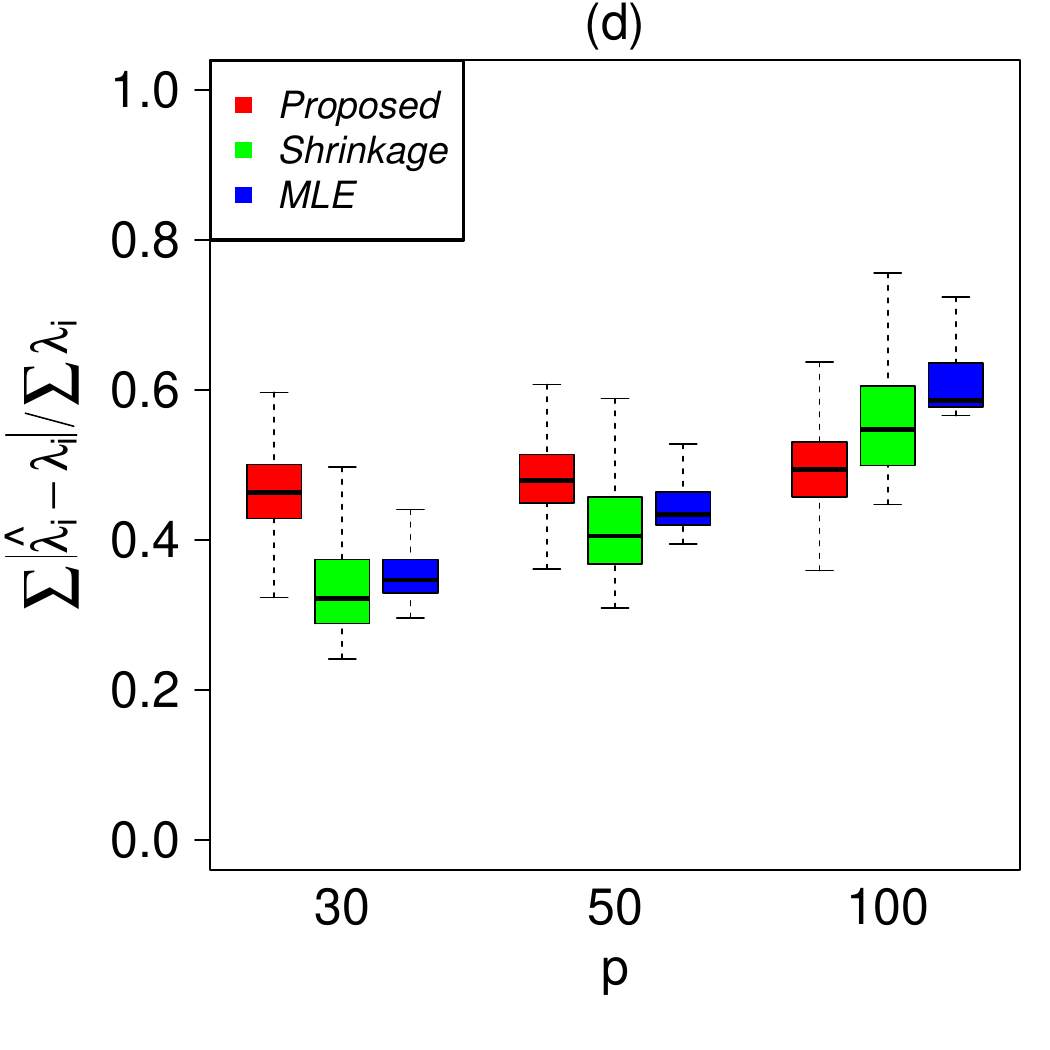}
\end{array}$
\end{center}
\begin{center}$
\begin{array}{r}
\includegraphics[scale=0.33]{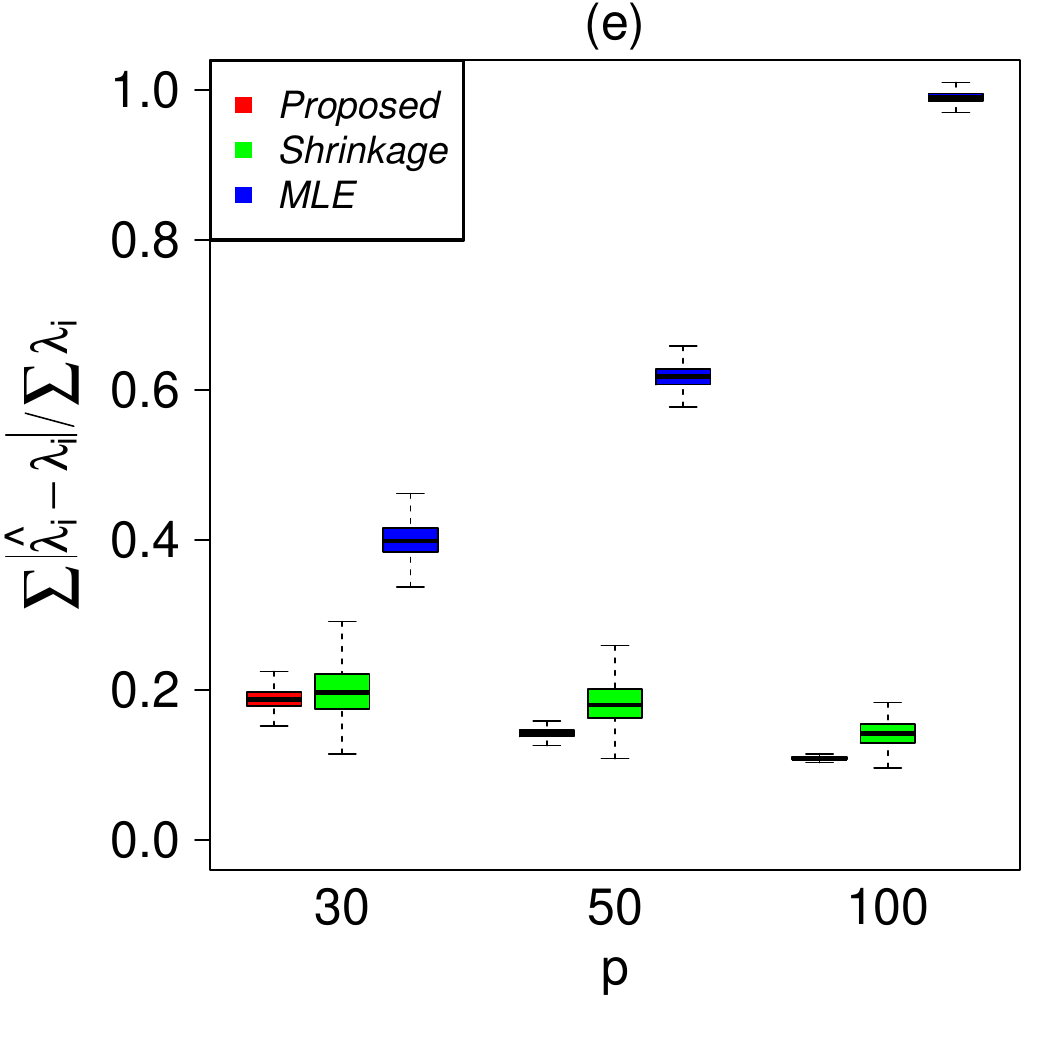}
\end{array}$
\end{center}
\caption{Distribution of the sum of absolute errors in estimated eigenvalues over 1000 simulations using maximum likelihood estimate, proposed method, and shrinkage estimate of \cite{schafer2005shrinkage} under different choices of covariance matrices: (a) and (b) true covariance is AR(1) with $t=0.5$, (c) and (d) true covariance is exchangeable with $t=0.5$ and (e) true covariance is random of structure with 30\% off-diagonal entries as non-zero. For the proposed method the target is: (a) AR(1), (b) misspecified as identity matrix, (c) exchangeable, (d) misspecified as identity matrix and (e) identity matrix. The data are generated from multivariate normal distribution with $n=50$ and $p \in\lbrace 30,50,100\rbrace$.}
\label{Fig1}
\end{figure}

The covariance matrix is estimated using the proposed method and shrinkage method of \cite{schafer2005shrinkage}. For the proposed method we estimate $t$ using Gaussian estimating equations as described in Section \ref{section3}. The sum of absolute errors in estimated eigenvalues averaged over 1000 simulated datasets are presented in Figure \ref{Fig1}. The eigenvalues of sample covariance are also plotted for comparison purpose.

From the simulation results, it is clear that whenever the target is correctly specified in case of both AR(1) and exchangeable covariance structures, the proposed method performs better than the shrinkage method as it maintains the smallest estimation error and is also much precise than the competing methods. Moreover, as $p$ increases the estimates obtained by using the proposed method become more accurate and precise. Its performance can become slightly weaker than the shrinkage method if the target is incorrectly specified as the identity matrix while the true covariance matrix is AR(1) or exchangeable. In case of random covariance structure when we use the identity matrix as a target, our proposed method also outperforms the shrinkage method in terms of accuracy and precision.

\section{Practical application} \label{section5}
In this section, we apply the proposed method to real world data of soil compaction profiles previously analysed by \citep{ullah2015regularised}. The data are from seven different ridges in the eastern Qilian Mountains, China, a rangeland habitat. Along each sunny ridge slope, samples are taken at three different heights. The heights are categorized as low, medium and high, so there are total 21 locations. At each of these 21 locations, measurements are taken at 18 consecutive depths (depths are our variables) from 2.5cm to 45cm with an interval of 2.5cm. So in summary, we have a total of 18 variables (depths) and 21 observations each of them can be categorised into one of the three hight levels that is low, medium and high.

\begin{figure}[H]
  \centering
\subfloat{\includegraphics[width=10cm]{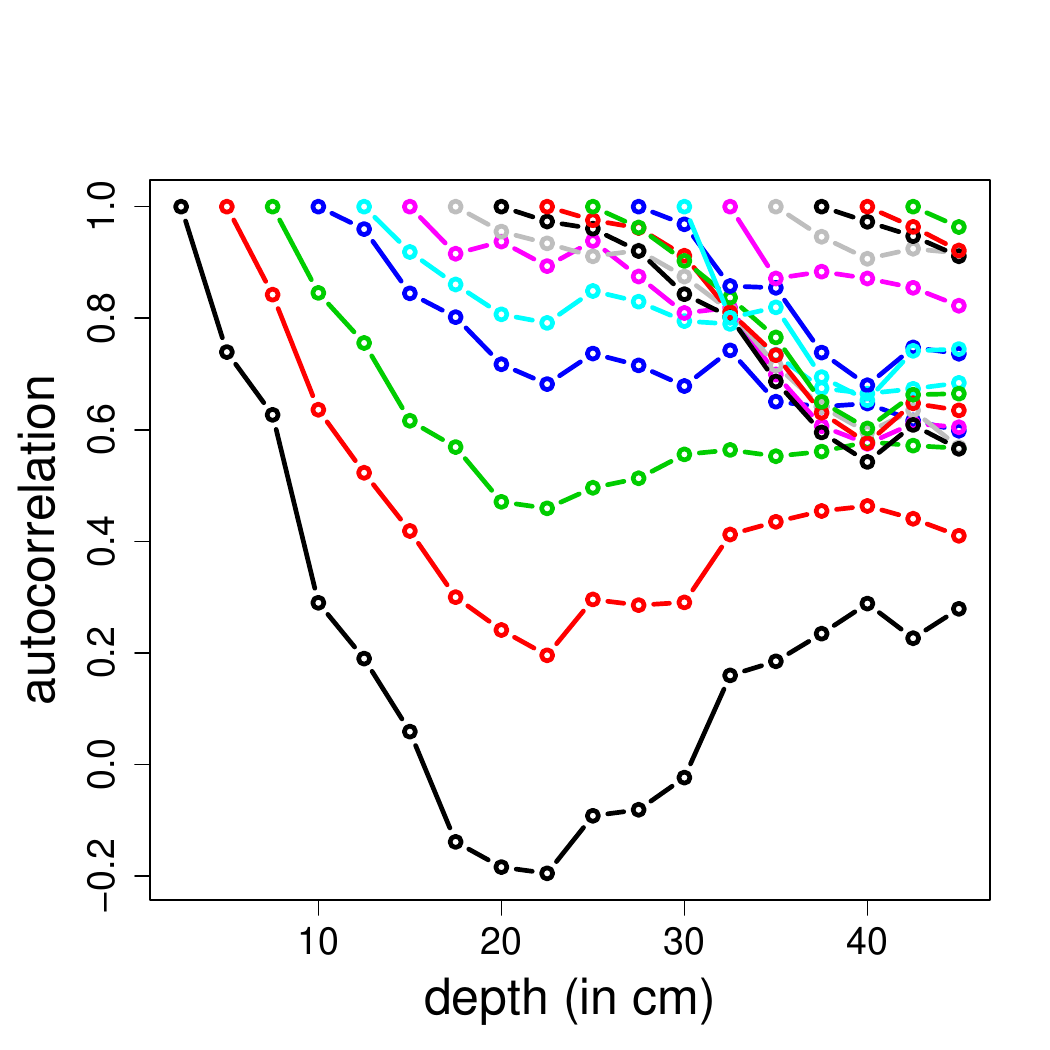}}\\
  \caption{Each sequence of connected points shows the correlation of measurements at a particular depth with those at greater depths. For each of the 18 depths (variables) we have 21 observations. The correlations are high for proximate depths, but decreases with increasing distance.}\label{autcorplot}
\end{figure}

\begin{figure}[H]
  \centering
\subfloat{\includegraphics[width=10cm]{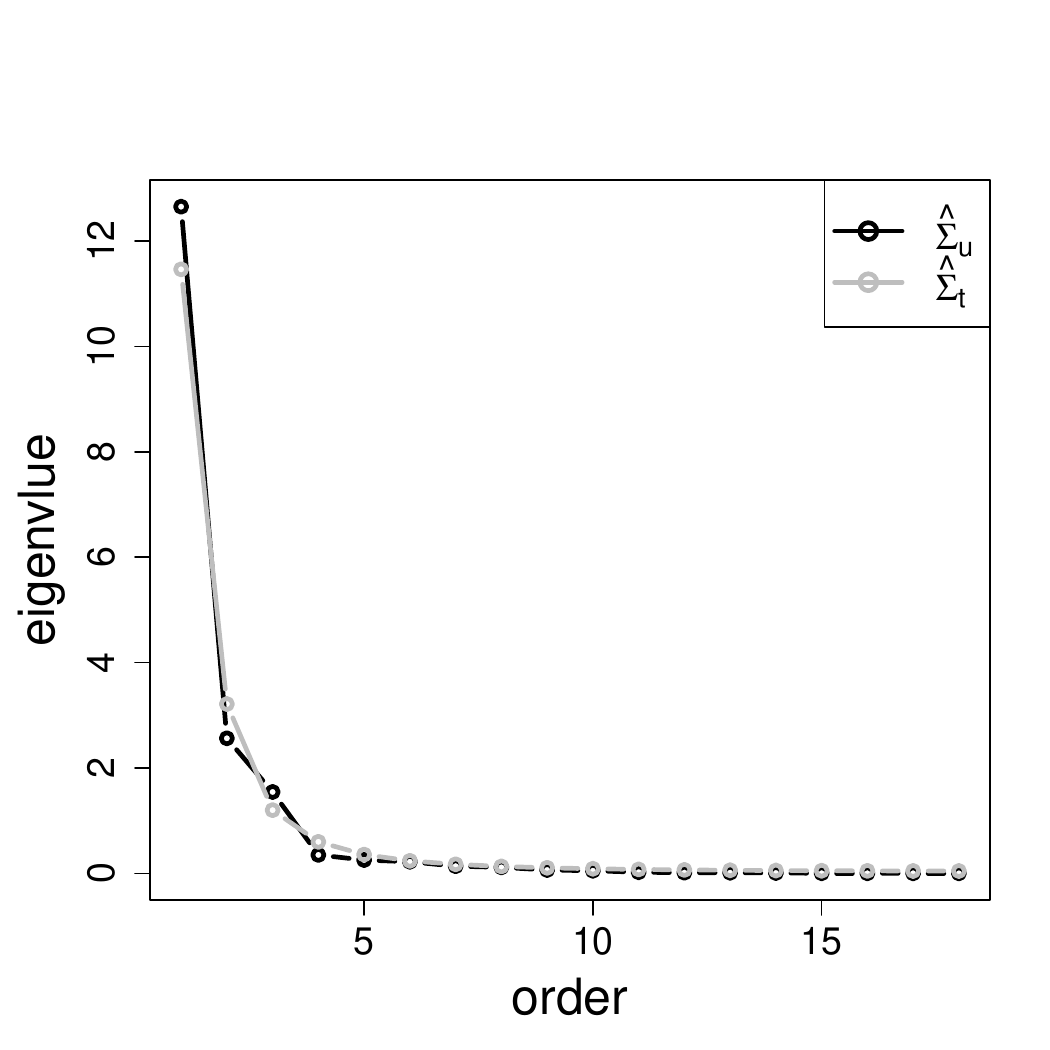}}\\
  \caption{Scree plots based on eigenvalues of $\widehat{\Sigma}_u$ and $\widehat{\Sigma}_t$.}\label{screeplot}
\end{figure}

To test the height effect (low, medium and high), we used multivariate analysis of variance (MANOVA) \citep{anderson1958introduction}. The group-centred observations for consecutive depths are highly correlated and the correlation decreases with increasing distance between depths suggesting correlation of an AR(1) structure (see Figure \ref{autcorplot} which is presented in \cite{ullah2015regularised} and reproduced here for convenience). We obtain the estimate of the parameter $t$ using the method described in Section \ref{section3} assuming an AR(1) structure, $\hat{t}=0.9162$. Figure \ref{screeplot} shows the eigenvalues of the sample correlation matrix of the group-centred observations in comparison to the eigenvalues from the estimated AR(1) structure with $\hat{t}=0.9162$, which further supports the perceived AR(1) structure. 

To compare the performance of the two methods in the three situations $n>p$, $n=p$ and $n<p$, we took random samples of different sizes (21, 18, 15, and 12) from 21 observations. The first set of 1000 samples each of size 21 was selected in such a way so that each of the three groups contribute seven observations with replacement, the second set of 1000 samples each of size 18 was selected in such a way so that each of the three groups contribute six observations with replacement, and so on. We fitted MANOVA model to each set of 1000 samples and summarized the significant results (at 0.05 level of significance) of fitted MANOVA models in Table \ref{Table}.   

The MANOVA model based on our proposed method and shrinkage method performed equally well in the $n>p$ case, that is, the mean difference was significant under both methods for all 1000 samples. However, the performance of the shrinkage method started to deteriorate sooner than the proposed method as we reduced the sample sizes. In the situation of $n=p$, the proposed method maintained the significance rate at 100\% while the significance rate for shrinkage method dropped to 97.8\%. Finally, in the situation of $n<p$   , the significance rate for the proposed method (99.9\% and 92.5\%, respectively, under the sample size 15 and 12) remained much higher than significance rate of the shrinkage method (92.5\% and 61.1\%, respectively, under the sample size 15 and 12).

\begin{table}[H] \centering 
\caption{Proportion of p-values less than 0.05. 1000 p-values are calculated by randomly selecting 7, 6, 5 and 4 samples (ridges) from each group using two different competing procedures.} 
  \label{Table} 
\begin{tabular}{l*{6}{c}r}
\hline \hline
No. of observations    &  Proposed method  & Shrinkage \\
\hline 
21 & 1.000 & 1.000  \\
18 & 1.000 & 0.978 \\
15 & 0.999 & 0.883 \\
12 & 0.925 & 0.611 \\
\hline
\end{tabular}
\end{table} 


\section{Conclusion and discussion} \label{section6}

In the high-dimensional setting, sample covariance exhibits some unfavourable characteristics. This is generally overcome by applying some regularization to the sample covariance, incorporating some additional information via shrinking the sample covariance towards a target. The most commonly used target is the identity matrix, probably because it satisfies some of the fundamental requirements of a good target. However, in some applications other structural targets may be preferred. 
In this paper, we have proposed to shrink the sample covariance matrix towards informative targets, namely the AR(1) and exchangeable covariance structures. These targets depend on an additional parameter that needs to be estimated, which we estimate by maximizing the log-likelihood function of the multivariate normal distribution.      
The regularized estimator depends on the penalty parameter, whose value needs to be chosen from within a range of possible values. We propose to choose the value of the penalty parameter by maximizing the log-likelihood function of the multivariate normal distribution. 
Being analytically simpler and computationally inexpensive, the proposed estimator is not only invertible but also well-conditioned. 

To evaluate the empirical performance of the proposed estimator, we conducted a simulation study. The simulation experiments showed the improved performance of the proposed estimator over the shrinkage estimator whenever the target is correctly specified in case of AR(1) and exchangeable covariance structures. However, its performance became slightly weaker than the shrinkage estimator when the target was incorrectly specified as AR(1) or exchangeable while the true covariance matrix was the identity matrix. In the case of random covariance structure our proposed estimator performed better than the shrinkage estimator. The proposed method was also tested on a real world dataset and we found that the proposed method maintained the significance rate (when embedded in MANOVA model) comparatively much better than the shrinkage method when the sample size was reduced.


\end{document}